\documentclass{article}

\usepackage{verbatim} 
\usepackage{float}

\usepackage{PRIMEarxiv}

\usepackage[utf8]{inputenc} 
\usepackage[T1]{fontenc}    
\usepackage{hyperref}       
\usepackage{url}            
\usepackage{booktabs}       
\usepackage{amsfonts}       
\usepackage{nicefrac}       
\usepackage{microtype}      
\usepackage{lipsum}
\usepackage{fancyhdr}       
\usepackage{graphicx}       

\usepackage{mathtools}
\usepackage{amsthm,amsmath,amssymb,amsfonts,exscale,latexsym,float,eucal}

\title{Fourier ptychography multi-parameter neural network with composite physical priori optimization}

\author{
  Delong Yang \\
  School of Optics and Photonics, Beijing Institute of Technology \\
  Beijing, China \\
   \And
  Shaohui Zhang*\\
  School of Optics and Photonics, Beijing Institute of Technology \\
  Beijing, China \\
  \texttt{*zhangshaohui@bit.edu.cn} \\
   \And
  Chuanjian Zheng \\
  School of Optics and Photonics, Beijing Institute of Technology \\
  Beijing, China\\
   \And
  GuoCheng Zhou \\
  School of Optics and Photonics, Beijing Institute of Technology \\
   \And
  Lei Cao \\
  School of Optics and Photonics, Beijing Institute of Technology \\
   \And
  Yao Hu \\
  School of Optics and Photonics, Beijing Institute of Technology \\
   \And
  Qun Hao*\\
  School of Optics and Photonics, Beijing Institute of Technology \\
  \texttt{*qhao@bit.edu.cn}
}

\begin{document}
\maketitle

\begin{abstract}
Fourier ptychography microscopy(FP) is a recently developed computational imaging approach for microscopic super-resolution imaging. By turning on each light-emitting-diode (LED) located on different position on the LED array sequentially and acquiring the corresponding images that contain different spatial frequency components, high spatial resolution and quantitative phase imaging can be achieved in the case of large field-of-view. Nevertheless, FPM has high requirements for the system construction and data acquisition processes, such as precise LEDs position, accurate focusing and appropriate exposure time, which brings many limitations to its practical applications. In this paper, inspired by artificial neural network, we propose a Fourier ptychography multi-parameter neural network (FPMN) with composite physical prior optimization.  A hybrid parameter determination strategy combining physical imaging model and data-driven network training is proposed to recover the multi layers of the network corresponding to different physical parameters, including sample complex function, system pupil function, defocus distance, LED array position deviation and illumination intensity fluctuation, etc. Among these parameters, LED array position deviation is recovered based on the features of brightfield to darkfield transition low-resolution images while the others are recovered in the process of training of the neural network. The feasibility and effectiveness of FPMN are verified through simulations and actual experiments. Therefore FPMN can evidently reduce the requirement for practical applications of FPM.
\end{abstract}

\section{Introduction}

In conventional microscopic systems, there is often a trade-off between field of view(FOV) and spatial resolution limited by the objective numerical aperture. The spatial resolution in a microscope is determined by numerical aperture (NA) of the objective and the illumination wavelength. Conventional method to solve the contradiction between large FOV and high resolution is to combine large NA objective and high-precision two-dimensional mechanical scanning, which increases the complexity and cost of the whole microscopic system simultaneously. Unlike the conventional way of relying entirely on hardware modification, Fourier ptychography microscopy (FPM), as one typical example of computational imaging methods, can achieve large FOV and high resolution at the same time with a cost-effective platform and without any mechanical scanning devides. Instead of scanning in the spatial domain, FPM scan and stitch the sample in the Fourier domain by illuminating it from different directions with a LED array\cite{zheng2013wide,song2019super,zheng2021concept,ou2013quantitative,konda2020fourier}. The prerequisite of FPM to successfully achieve resolution enhancement is the phase retrieval of each sub-aperture in frequency domain, and a certain degree of sub-aperture overlap is the key to ensure that phase retrieval can be achieved. In fact, the information redundancy caused by the sub-spectrum overlap can not only ensure the recovery of sample complex information, but also some system parameters or experimental parameters, such as coherent transfer function of system, LED array position deviation, defocus distance of sample and illumination intensity fluctuation from different LEDs(hereafter refer to as other factors)\cite{li2021biopsy}.

There are many method using characteristics of data redundancy to recover other factors, Ou, Xiaoze et al proposed Embedded pupil function recovery method to recover the pupil function of the system.\cite{ou2014embedded}, Sun Jiasong et al. and Eckert Regina et al. proposed method to recover LED array position deviation\cite{sun2016efficient,eckert2018efficient,dwivedi2018lateral}. However, the conventional optimization algorithm for reconstruction needs to mathematically find analytical differentiation of captured low-resolution images (LR images) on the target to be recovered\cite{zuo2016adaptive,konijnenberg2017introduction}. However, it is diffcult to find the analytical differentiation of LR images on other factors, such as defocus distance of sample and illumination intensity fluctuation of LEDs, which also makes it difficult for conventional algorithms to incorporate the other factors into optimization variables. And because the other factors are coupled in the imaging process, it is difficult to optimize multiple parameters simultaneously\cite{zhang2019pgnn}. 
Due to the difficulty of finding the analytical differentiation of some factors at present stage\cite{jurling2014applications,kreutz2009complex}, we turn to numerical differential method\cite{barbastathis2019use}. The training process of neural network\cite{lecun2015deep} is optimizing the parameters of network according to the numerical differentiation. And there is no need to find the analytical differentiation of loss on network parameters\cite{wang2020phase,thanh2018deep}. As the structure of network is established, it can automatically find the numerical differentiation. Moreover, thanks to rapid development of deep learning, many loss functions can be constructed to improve the speed of solving optimization problems and achieve global optimum\cite{ruder2016overview,kingma2014adam}. When solving inverse problems through neural network, we can make full use of neural network training tools to realize optimal solution of multi-parameters\cite{lin2018all}. In 2018, SHAOWEI JIANG et al. first proposed to model FPM forward propagation process with convolutional neural network\cite{jiang2018solving}. They get the complex object information through training process of network.However, their method take the process of lightwave through microscope is ideal, and no other factors are taken into account. 

To improve the robustness of neural network reconstruction algorithm, and achieve multi-parameter reconstruction, we model the other factors in the system as network layers. Since the convolution neural network will lead to rapid growth in computing volume with the size of image increasing\cite{lecun2015deep}, we use element-by-element multiplication between layers representing different factors to increase the computation efficiency. But at the same time, the increase of parameters to be optimzed will cause pathological degree of problem to increase\cite{tikhonov1963solution}. Zhao, Ming et al. have tried modeling the LED array position deviation into neural network, but the optimization process is very time-consuming\cite{zhao2021neural}. Since using physical means to reduce the morbidity of inverse problems is a very effective method\cite{zhang2021fast,zhou2020adaptive,kalita2021single}, we have reduced the pathologicality of phase recovery problems through inserting a wedge angle in front of the microscope to imporve the speed and quality of reconstruction\cite{zhang2021asymmetric}.In this paper, in order to reduce the pathological degree of reconstruction problem, we turn to find physical method to correct LED array position deviation. We choose four brightfield to darkfield transition LR images which located on orthogonal direction, using boundary of bright-field to dark-field transition on the LR images to calculate LED array position deviation inspired by the method proposed in reference\cite{zheng2021concept}. We named the neural network proposed in this paper for FPM reconstruction algorithm as Fourier Ptychographic Multiparamater Net(FPMN), the physics method for correcting LED array position deviation as Array Correction Fourier ptychography(ACFP). In FPMN, the lightwave propagation process is modeled as fixed parameters of network layer (discrete reverse fourier transformation). We implement model construction with deep learning framework pytorch, using automatic differentiation of pytorch to optimize network parameters. 

This paper is structutred as follows. The principle of standard FPM framework and system setup are presented in section 2.1. The working principle and structure of FPMN are presented in section 2.2$\sim$2.3. The definition of LED array position deviation and the principle of ACFP are presented in 2.4. In section 3, we conduct simulation experiments with different defocus distance and mixed deviations(defocus aberration and LED array position deviation), comparing the reconstruction quality achieved through EPRY and our method. The USAF chart and biological samples are used to demonstrate the effectiveness of FPMN in section 4.1$\sim$4.2. We disscus the influence of different pupil function model method on reconstruction quality in 4.3. Conclusions and disscusions are given in Section 5. 

\section{Principles}
\subsection{FPM principle}
\begin{figure}[H]
	\centering
	\includegraphics[scale=0.45]{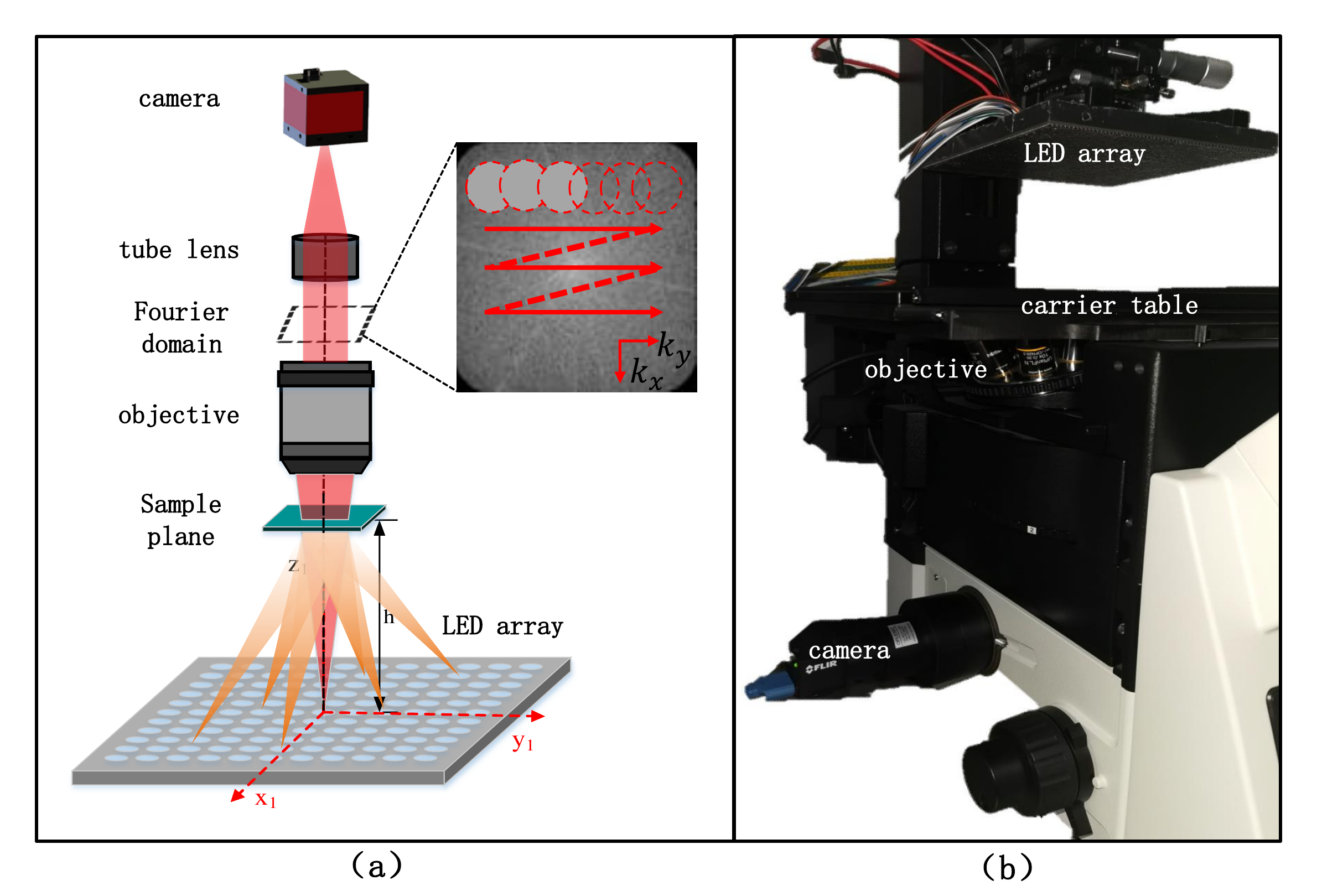}
	\caption{Imaging principle of FPM and actual system}
\end{figure}
Unlike conventional imaging, which relies more on hardware systems, computational imaging refers to joint optimization of physical models, imaging pipelines, and inverse problem algorithms\cite{kellman2020memory}. Therefore, clear physical model and accurate model parameters during imaging process are essential for the success of FPM, As a representative computational imaging method shown in Fig.1(a), a typical FPM system consists of a LED array providing angular varied illuminations, an microscopic imaging module collecting the diffracted light emitted from the sample, and a image sensor to receive and record the two-dimensional intensity images. For a small sample segment, whose size is much smaller than the distance between the LED array and sample, the illumination wave can be approximately treated as a parallel plane wave. Thus, according to the Fourier optics, illumainating the sample by an oblique plane wave with a wavevector (kx,ky) is equivalent to shifting the sensor of the sample's spectrum by (kx,ky) in the Fourier domain. 
The microscopic objective acts as the combination of a Fourier transformer and a filter in Fourier domain. Therefore, the forward light field propagating and imaging process of FPM can be expressed as
\begin{align}
\mathbf{I}_{n}(x, y) = |\mathbf{F}^{-1}\{\mathbf{F}\{\mathbf{t}(x, y)\}\cdot\mathbf{P}(k_{x}, k_{y})\}|^2
\end{align}
where $\mathbf{t}(x, y) = s(x, y) \cdot e^{i(x k_{xn} + y k_{yn})}$ denotes the exit wave distribution of the sample $s(x, y)$ that is illuminated by an oblique illumination with a wavevector $(k_{xn}, k_{yn})$. '$\mathbf{F}$' and '$\mathbf{F}^{-1}$' indicate the Fourier and inverse Fourier transform respectively. $\mathbf{P}(k_{x}, k_{y})$ is the pupil function of the objective,$(x, y)$ is the 2D spatial coordinates in the spatial domain and $(k_{x}, k_{y})$ is the corresponding spatial frequencies in the frequency domain. 
$I_{n}(x, y)$ is the intensity image acquired by the camera. 

\subsection{The principle and structure of Fourier ptychography multi-parameters neural network}

According to the Eq.(1), the forward propogation of light field in FPM can be modeled as an element-wise neural network. In order to simplify the network and reduce the amount of computation, we rewrite Eq.(1) as Eq.(2) to directly model the sample in frequency domain. 
Since complex differentiation is not supported in the advanced Neural network framework Pytorch, we model the sample complex function as two sub-channel layers. We model the the complex pupil function $\mathbf{P}(k_{x},k_{y})$ in Eq.(2) corresponding network layer with the top ten Zernike coefficients, a classic two-dimensional phase distribution representation approach.This method of incorporating calssical physical models into the network can evidently reduce the complexity of the network and the ill-conditioning degree of the network training. The computational relationship between layers can be express as
\begin{align}
I_{n}(&x, y) = |\mathbf{F}^{-1}(\hat{\mathbf{O}}(k-k_{n}) \cdot \mathbf{P}(k_{x}, k_{y}))|^{2}  ,n = 1, 2, \cdots N^{2}\\
\hat{\mathbf{O}}(k) &=\ \hat{\mathbf{O}}_{r}(k) + \hat{\mathbf{O}}_{i}(k), \quad 
\mathbf{P}(k_{x}, k_{y}) =  T_{Zernike}(z_{1}, z_{2}, \ \cdots , \ z_{10})
\end{align}
\begin{align}
\left\{
\begin{array}{lr}
\mathbf{E}_{nr} = \hat{\mathbf{O}}_{r}(k-k_{n})\cdot cos(\mathbf{P}(k_{x}, k_{y})) - \hat{\mathbf{O}}_{i}(k-k_{n})\cdot sin(\mathbf{P}(k_{x}, k_{y})) \ \\
\mathbf{E}_{ni} = \hat{\mathbf{O}}_{r}(k-k_{n})\cdot sin(\mathbf{P}(k_{x}, k_{y})) + \hat{\mathbf{O}}_{i}(k-k_{n})\cdot cos(\mathbf{P}(k_{x}, k_{y})) ,n = 1, 2\ \cdots , N^{2}
\end{array}
\right.
\end{align}
\begin{align}
\mathbf{I}_{n}(x, y) = |\mathbf{F}^{-1}\{\mathbf{E}_{nr} + \mathbf{E}_{ni}\}|^2 ,n = 1, 2\ \cdots , \ N^{2}
\end{align}
where $\mathbf{O}(k)$ denotes the Fourier transform of the object function,$k_{n}$ denotes the illumination vector, $\mathbf{P}(k_{x},k_{y})$ represents the pupil function(radian of phase modulation in Eq.(3)), $N$ represents the number of rows or columns of LEDs on LED array, $\hat{\mathbf{O}}_{r}(k)$ and $\hat{\mathbf{O}}_{i}(k)$ are the real part and imaginary part respectively, $\mathbf{E}_{nr}$ and $\mathbf{E}_{ni}$ are the real part and imaginary part of exit wave distribution through the pupil. The inverse Fourier transformation in Eq.(5) are then modeled as a four-channel layer with fixed parameters. Then the light field impinging on the image detector surface can be expressed as
\begin{align}
\centering
\mathbf{F}^{-1}(\mathbf{E}_{n}) = \mathbf{G}_{a} \cdot \mathbf{E}_{n} \cdot \mathbf{G}_{b} \rightarrow
\mathbf{F}^{-1}(\mathbf{E}_{n}) = (\mathbf{G}_{ar}+\mathbf{G}_{ai}) \cdot (\mathbf{E}_{nr}+\mathbf{E}_{ni}) \cdot (\mathbf{G}_{br}+\mathbf{G}_{bi})
\end{align}
where $\mathbf{G}_{ar}$ and $\mathbf{G}_{ai}$ denote the real part and imainary part of $\mathbf{G}_{a}$, $\mathbf{G}_{br}$ and $\mathbf{G}_{bi}$ denote the real part and imainary part of $\mathbf{G}_{b}$, $\mathbf{G}_{a}$ and $\mathbf{G}_{b}$ can be express as
\begin{align}
\mathbf{G_{a}} = 
\begin{pmatrix}
e^{-j2\pi\frac{0\cdot0}{M}} & e^{-j2\pi\frac{0\cdot1}{M}} & \cdots & e^{-j2\pi\frac{0\cdot(M-1)}{M}} \\
e^{-j2\pi\frac{1\cdot0}{M}} & e^{-j2\pi\frac{1\cdot1}{M}} & \cdots & e^{-j2\pi\frac{1\cdot(M-1)}{M}} \\
\vdots                        & \vdots                      & \ddots & \vdots                        \\
e^{-j2\pi\frac{(M-1)\cdot0}{M}}&e^{-j2\pi\frac{(M-1)\cdot1}{M}}&\cdots&e^{-j2\pi\frac{(M-1)\cdot(M-1)}{M}}
\end{pmatrix}
\end{align}
\begin{align}
\mathbf{G_{b}} = 
\begin{pmatrix}
e^{-j2\pi\frac{0\cdot0}{N}} & e^{-j2\pi\frac{0\cdot1}{N}} & \cdots & e^{-j2\pi\frac{0\cdot(N-1)}{N}} \\
e^{-j2\pi\frac{1\cdot0}{N}} & e^{-j2\pi\frac{1\cdot1}{N}} & \cdots & e^{-j2\pi\frac{1\cdot(N-1)}{N}} \\
\vdots                        & \vdots                      & \ddots & \vdots                        \\
e^{-j2\pi\frac{(N-1)\cdot0}{N}}&e^{-j2\pi\frac{(N-1)\cdot1}{N}}&\cdots&e^{-j2\pi\frac{(N-1)\cdot(N-1)}{N}}
\end{pmatrix}
\end{align}

Since we have split the real-imaginary part of the complex light field, we break $\mathbf{G_{a}}$ down into $\mathbf{G_{ar}}$ and $\mathbf{G_{ai}}$ in Eq.(6) according to the Euler formula. Similarly, $\mathbf{G_{b}}$ can also be decomposed into $\mathbf{G_{br}}$ and $\mathbf{G_{bi}}$. 
In addition to these hidden layers, we model the sequential angular varied illumination plane waves as the input layer, and the two-dimensional intensity images as the output layer, respectively. Up to now, as shown in Fig.2(a), the neural network corresponding to the ideal FPM imaging process has been constructed successfully. 
\begin{figure}[H]
	\centering
	\includegraphics[scale=0.28]{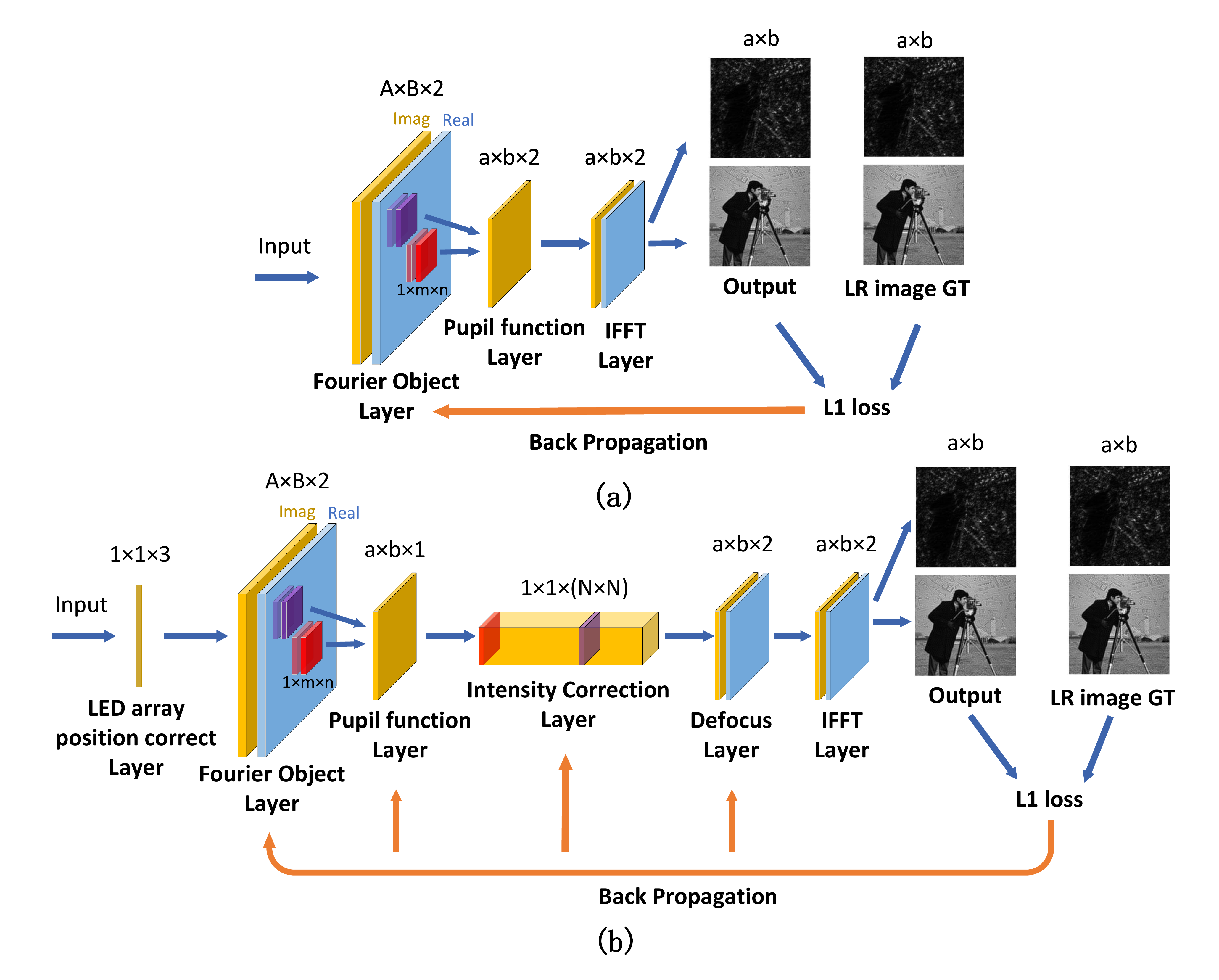}
	\caption{(a)Structure of FPN(Fourier Ptychography Network) with ideal FPM imaging process.(b)Structure of FPMN}
\end{figure}

As mentioned previously,the HR complex function of the sample can be recovered through by training of the neural network,with the help of the advanced various network optimizers. There are several ways to construct the loss functions optimizing the neural network, such as L0,L1, and L2, etc. We choose L1-norm as the loss function in our approach\cite{jiang2018solving}, as shown in the following equation. 
\begin{align}
\mathbf{loss} = diff(\mathbf{I}^{gt}_{n} - \mathbf{I}^{predict}_{n}) = \sum_{n=1}^{N}|\mathbf{I}^{gt}_{n} - \mathbf{I}^{predict}_{n}|\quad(n = 1, 2 \ \cdots \, N^{2})
\end{align}

It is worth noting that the accurate position of each sub-spectrum needs to be known when updating the HR spectrum for high recovery quality. In other words, the illumination wave vectors determined by the position of each LED element and the distance between LED array and sample need to be precisely known. The LED array position deviation is defined in Fig.3.Since the pitch and yaw of the LED board are easily zeroed with a spirit level, we only take the horizontal movement and rotation angle about the optical axis into consideration. Where $\Delta x$ and $\Delta y$ in Fig.3 respectively denotes the translation of LED array in two orthogonal directions,$\theta$ denotes the rotation of the LED array. 
\begin{figure}[H]
	\centering
	\includegraphics[scale=0.28]{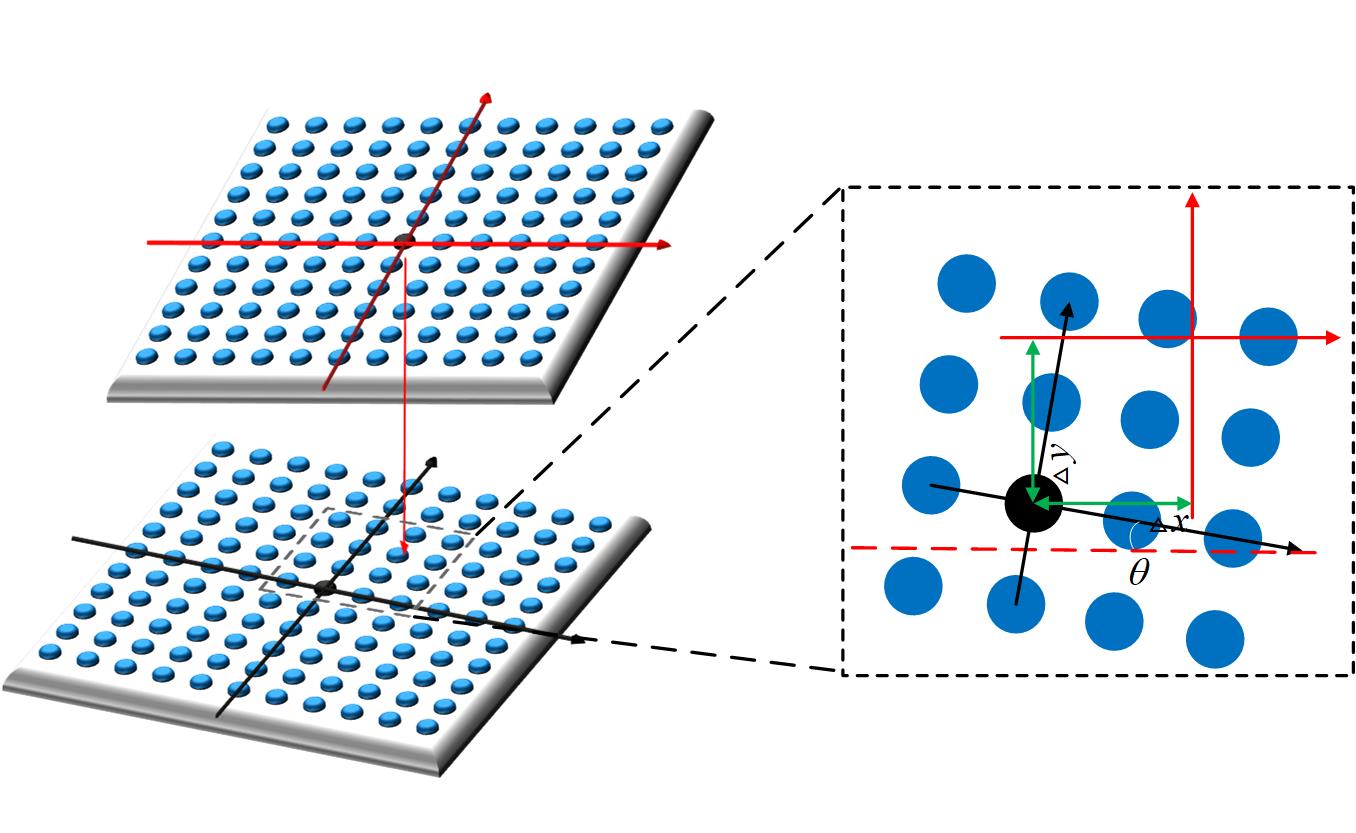}
	\caption{The rotation and translation of LED array in the horizontal plane}
\end{figure}
Unlike the ideal FPM model that has no errors, there are several critical system parameter errors in the system construction and raw data acquisition process. As shown in Fig.2(b), we add three dominal system errors, the LED array position deviation, the sample defocus distance, and the illumination intensity fluctuation into the ideal FPM forward network. In the original FPM model, we suppose each LED element is a point source that emits quasi-monochromatic light, and the whole LED array should be aligned. Nevertheless, the positional deviation of the LED board is unavoidable during the process of constructing or modifying FPM systems. So we add a LED array position correct layer, we use $\Delta x, \Delta y, \theta$ in Fig.3 as the parameters in this layer and the corrected sub-aperture position can be express as
\begin{align}
x_{n}^{correct} &= cos(\theta) \cdot x_{n} + \Delta x, \
y_{n}^{correct} = sin(\theta) \cdot y_{n} + \Delta y, \ n = 1,2 \ \cdots \,N^{2} \\
\theta_{xn}^{correct} &= arctan(\frac{x_{n}^{correct}}{h}), \
\theta_{yn}^{correct} = arctan(\frac{y_{n}^{correct}}{h}), \ n = 1,2\ \cdots \, N^{2} \\
k_{n} = (k_{xn}&, k_{yn}), \ k_{xn} = 2\pi \frac{sin(\theta_{xn}^{correct})}{\lambda}, \ k_{yn} = 2\pi \frac{sin(\theta_{yn}^{correct})}{\lambda} , \ n = 1,2 \ \cdots \,N^{2}
\end{align}

The ideal FPM model assume that the sample is a flat thin simple in focus, while the non-planar distribution and improper focusing will cause different defocus distances for parts or the whole sample. Therefore, in FP imaging process, we should take the defocus distance of sample into account, modeling effect of defocus in FPMN, so that the recovery of the complex object can be free from the defocus aberration. Accroding to the defocus distance of sample, the distribution of the light field after free-space propogation can be expressed as
\begin{align}
\mathbf{E}_{1}(x, y) \ = \ \mathbf{F}^{-1}\{\mathbf{F}\{\mathbf{E}_{0}(x, y)\}\cdot \mathbf{H}(k_{x}, k_{y}, z)\} \\
\mathbf{H}(k_{x}, k_{y}, z) = exp(j \frac{2 \pi}{\lambda} \cdot z \cdot \sqrt{1-k_{x}^{2}-k_{y}^2})
\end{align}
where $\mathbf{E}_{0}(x, y)$ is a known complex light field in a focusing plane, $E_{1}(x, y)$ is the field in a plane with a distance of $z$. Accroding to Eq.(10), we model the free-space propogation as an defocus layer in FPMN. We set defocus distance z as the parameters to be optimized in defocus layer. The defocus layer is set behind the pupil function layer and before the IFFT fixed parameter layer. Similarly, the defocus parameter can also be recovered by training the FPMN simultaneously. Since pytorch does not support complex differentiation, free-space propogation for light field in FPMN with defocus taken into consideration can be expressed as
\begin{align}
\mathbf{E}_{nr}^{d}(x, y) \ = \ \mathbf{F}^{-1}\{\mathbf{F}\{{\mathbf{E}}_{nr}(x, y)\}\cdot \mathbf{H}_{r}(k_{x}, k_{y}, z) - \mathbf{F}\{\mathbf{E}_{ni}(x, y)\}\cdot \mathbf{H}_{i}(k_{x}, k_{y}, z)\} \\
\mathbf{E}_{ni}^{d}(x, y) \ = \ \mathbf{F}^{-1}\{\mathbf{F}\{\mathbf{E}_{nr}(x, y)\}\cdot \mathbf{H}_{i}(k_{x}, k_{y}, z) - \mathbf{F}\{\mathbf{E}_{ni}(x, y)\}\cdot \mathbf{H}_{r}(k_{x}, k_{y}, z)\}
\end{align}
where $E_{nr}^{d}$, $E_{ni}^{d}$ denotes real and imaginary part of the light field passing through defocus layer. $H_{r}$,$H_{i}$ denotes real and imaginary part of $H(k_{x},k_{y},z)$. 

In FPM system, there are inevitable illumination intensity fluctuation from different angle LEDs, Due to manufacturing limitations, there will be differences in the luminous brightness of LEDs in different positions. According to the lighting model in FPM, the LED illumination distance and illumination angle of differnet LEDs are different. Under the influence of above mentioned factors, it is difficult to explicitly calculate the intensity fluctuation of different angle LEDs. 
We think that further correction of intensity fluctuation can improve recovery quality of the complex sample. We compensate intensity deviation with intensity correction coefficient $\gamma$. After adding defocus correction, intensity correction, forward propagation process of FPMN can be expressed as
\begin{align}
\mathbf{I}_{n}(x, y) = |\mathbf{F}^{-1}(\gamma_{n} \cdot \hat{\mathbf{O}}(k-k_{n}) \cdot\mathbf{P}(k_{x}, k_{y}) \cdot \mathbf{H}(k_{x}, k_{y}, z)|^2 \quad (\gamma_{n} = \gamma_{1}, \gamma_{2}, \ \cdots \ \gamma_{N})
\end{align}

We model the process of intensity correction as FPMN network layer, which has $N^{2}$ parameters to optimize, each parameter corresponds to one LED illumination intensity fluctuation. The structure of the complete FPMN with all these parameters taken into consideration is illustrated in Fig.2(b). 

\subsection{LED array correction method(ACFP) through physics model}

Since the optimization of the LED array position deviation in FPMN will be very time-consuming\cite{zhao2021neural}. In this subsection, we proposed a correction method ACFP to correct the LED array position deviation through physics model. 
\begin{figure}[H]
	\centering
	\includegraphics[scale=0.11]{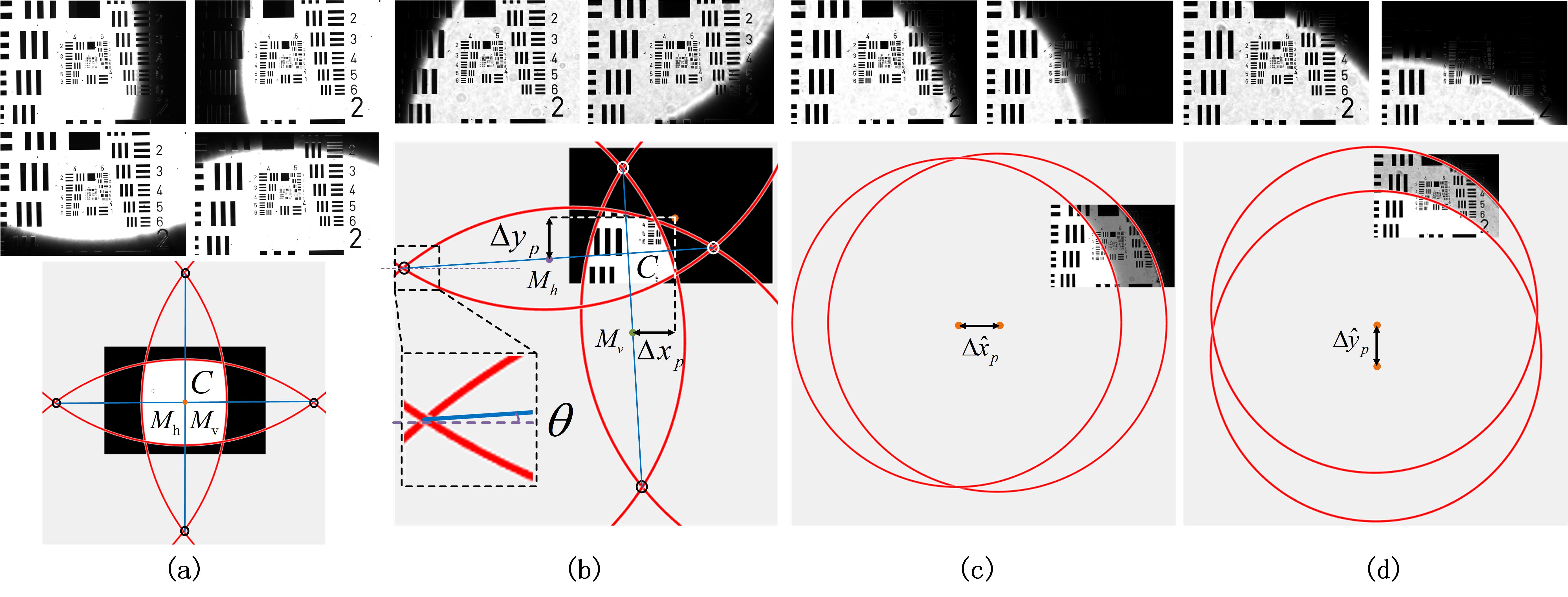}
	\caption{(a)LR images illuminated by LED in top,bottom,right and down of LED array center when there is no LED array position deviation. Fusiform zone formed by fitted circle. (b)(c)(d)There are translation and rotation on the LED array. (b)LR images illuminated by LED in left and bottom of LED array center. Fusiform zone formed by fitted circle. (c)LR images illuminated by two adjacent LEDs in the right of LED array center and fitted circle. (d)LR images illuminated by two adjacent LEDs in the top of LED array center and fitted circle. }
\end{figure}
As shown in Fig.4(a) and Fig.4(b), the bright-field to dark-field transition feature will appear in the image under some specific oblique angle illumination. The circular transition feature is caused by the circular shape of the circular pupil function of the imaging system. Since the relationship between the illumination direction and the bright-field image position is linear, we propose a framework to calculate the LED pose parameters according to the position characteristics of the bright-field image positions. Fig.4(a-d) indicate four LR images corresponding to four symmetrical LED illumination with known serial numbers. Symmetrical circles can form a fusiform zone whose location can be used for the recovery of the LED pose correctly. For convenience, we fit a bright-field circle by using the arc-shaped bright-field to dark-field transition boundary line, and use the center of the circle to indicate the position of the bright-field area. 

As shown in Fig.4, $M_{v}$ denotes the midpoint of the upper vertices and lower vertices. $M_{h}$ denotes the midpoint of the left vertices and right vertices. $C$ denotes the center of the full field of view LR image. When there is no LED array position deviation,as shown in Fig.4(a), $M_{v}$,$M_{h}$,$C$ coincident with one point. When the position of LED array is inaccurate, as shown in Fig.4(b), the deviation of $M_{h}$ and $C$ in y direction, the deviation of $M_{v}$ and $C$ in x direction, are respectively related to the LED array position deviation in y direction and x direction. 

Subsequently, we calibrate the scaling factor between the translation of the bright-field image and the LED array horizontal deviation with the help of bright-field lateral offset corresponding two adjacent LED units. As shown in Fig.4(c) and Fig.4(d), the pixel offset $\Delta\hat{x}_{p}$ and $\Delta\hat{y}_{p}$ correspond to the LED spacing, so $\Delta x_{p}$ and $\Delta y_{p}$ correspond to the translation of LED array can be express as
\begin{align}
\Delta x = \frac{\Delta x_{p}}{\Delta\hat{x}_{p}} \times P\\
\Delta y = \frac{\Delta y_{p}}{\Delta\hat{y}_{p}} \times P
\end{align}
where P denotes the LED spacing in LED array, $\theta$ in Fig.4(c) can directly represent the rotation angle of LED array. 

\subsection{The training process of FPMN}
\begin{figure}[H]
	\centering
	\includegraphics[scale=0.3]{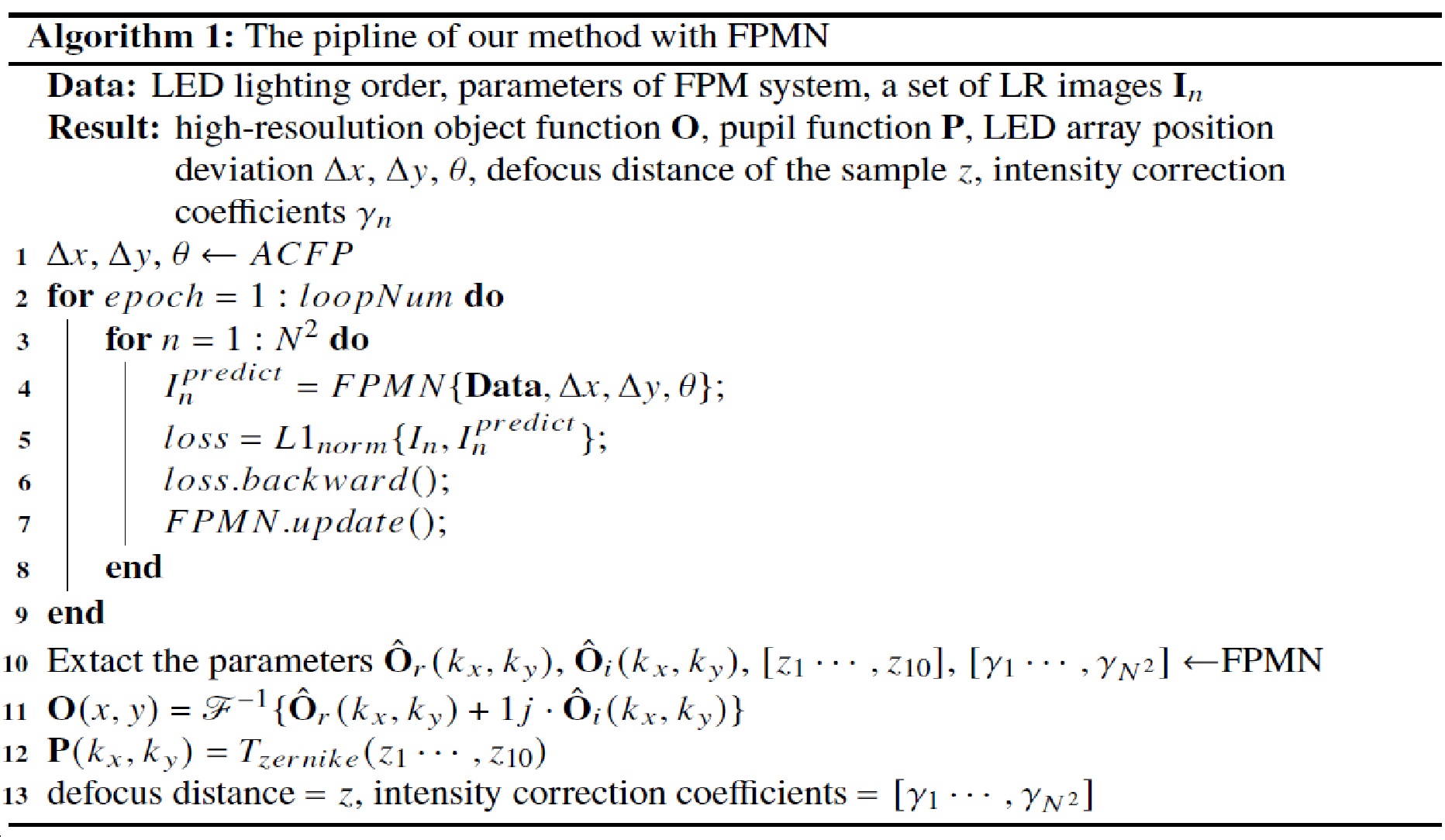}
	\caption{Algorithm outline of our proposed method.}
\end{figure}
The framework of multi-parameters reconstruction algorithm of the proposed method is shown in Fig.5. We choose four brightfield to darkfield LR images in two orthogonal directions to calculate the LED array position deviation by ACFP. The LED array position deviation is used as a supplementary input to FPMN. Initialize $\hat{O}_{x, y}=Ae^{j\varphi}$ as the complex sample parameters, with using the intensity distribution of central LR image as $A$ and $\varphi$. Initialized pupil function without aberration, defocus distance with zero, intensity correction coefficients with ones. 

\section{Simulations}
\begin{figure}[H]
	\centering
	\includegraphics[scale=0.15]{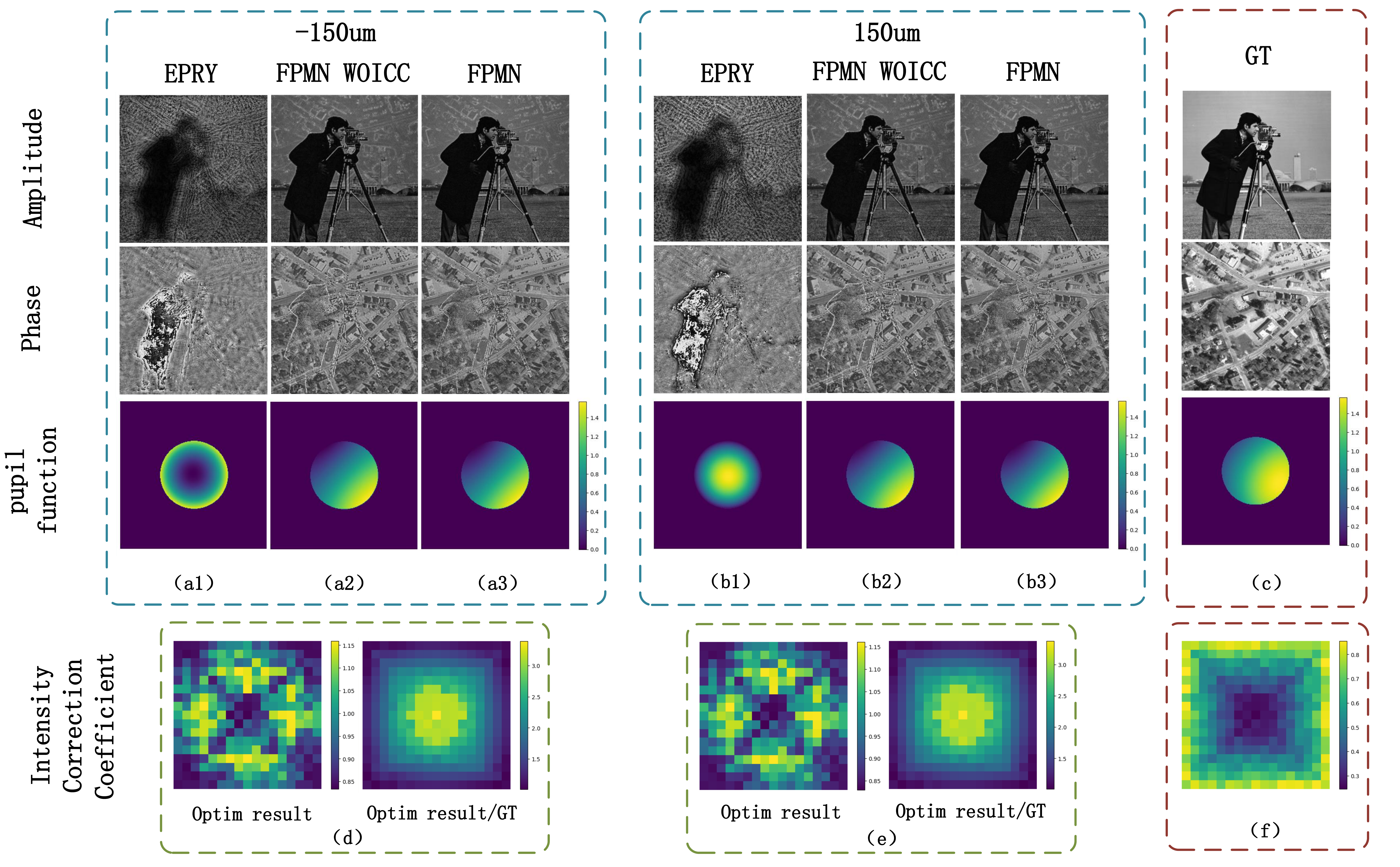}
	\caption{a(1), b(1)Results of EPRY in defocus state. a(2), b(2)Results of FPMN without ICC(intensity correction coefficients) in defocus state. a(3), b(3)Results of complete FPMN in defocus state. (c)Groundtruth of the sample and pupil function. The left of (d), (e) of ICC and ICC divide groundtruth. (f)Groundtruth of ICC. }
\end{figure}
\begin{figure}[H]
	\centering
	\includegraphics[scale=0.15]{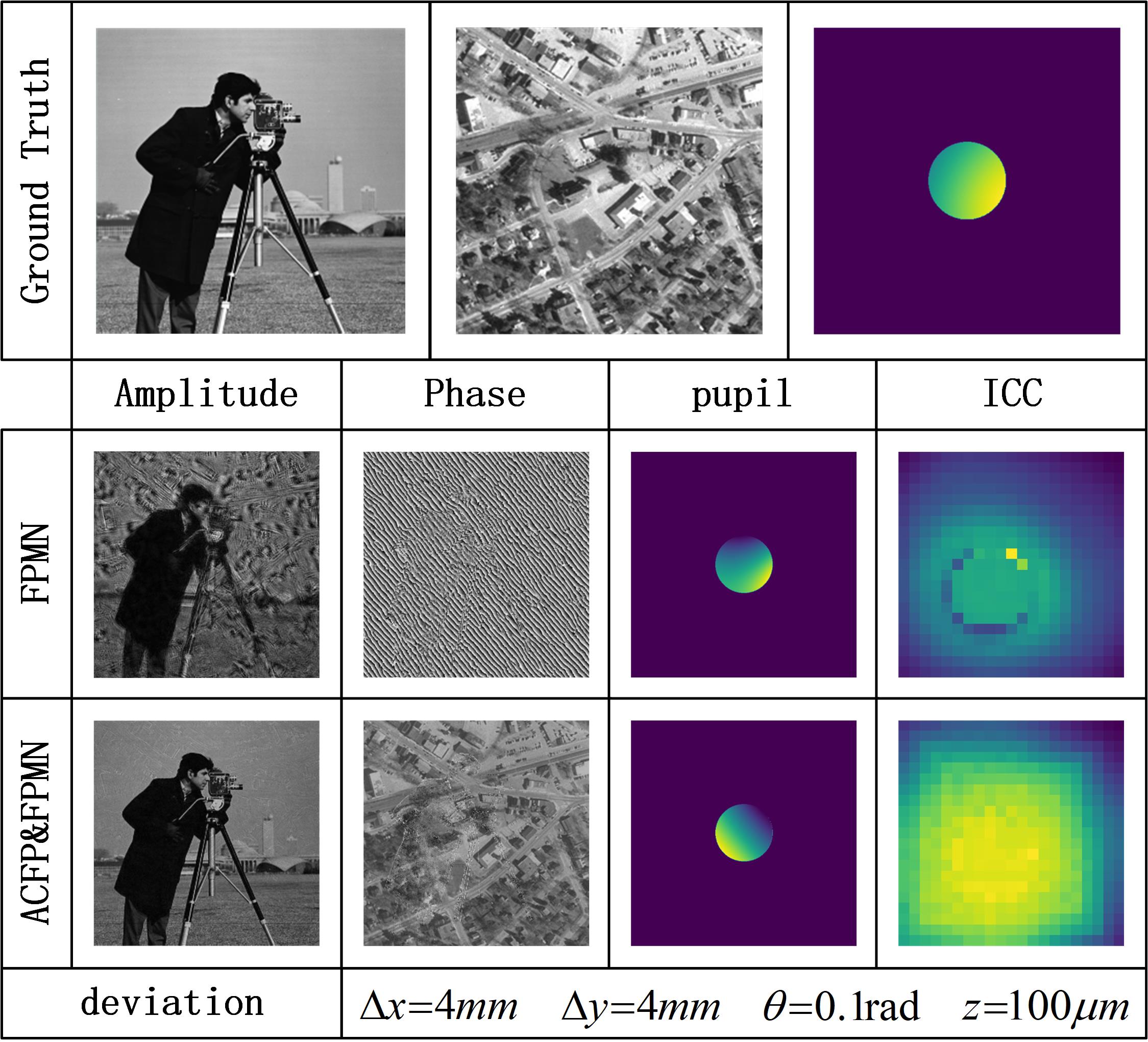}
	\caption{simulation results with defocus aberration and LED array position deviation}
\end{figure}
The simulation parameters were chosen to realistically model an FPM platform, with an incident illumination wavelength of 632.8nm, an image sensor with pixel size of 2.4mm, and an 4X objective with NA of 0.1. We simulated the use of the central $15\times15$ LEDs in the array placed 92mm beneath the sample, and the distance between adjacent LEDs is 2.5mm. The raw low-resolution data is limited to a small region with only $256\times256$ pixel resolution, and the final high-resolution complex field with $1024\times1024$ pixel is recovered by different approaches. We compared the results of our method (FPMN) and EPRY at different defocus state in Fig.6. And the result has improved after we activate the optimization of the intensity correction coefficients layer. As we can see in (d), (e) of Fig.6 that the tendency of the zero diffraction light to the secondary diffraction light is gradually decreasing. 

To verify the robustness for mixed diviation(defocus aberration and LED array position deviation), we also conduct a simulation experiment. As shown in Fig.7, amplitude and phase of reconstruction object were terrible when there are LED array position deviation in the system, but we can see the trend of translation in the recover result of intensity correction coefficients(ICC). As a comparation, when we obtain the LED array position deviation through ACFP, and enter it to LED array position correction layer of the FPMN, FPMN can still obtain high-quality reconstructed intensity image and phase image even with such large positional deviations. 

Afted introducing the different defocus aberration and intensity correction coefficients(adding base on exposure time and random intensity error) into the imaging process of simulation experiment, we plot the SSIM curve to compare the results with different method in Fig.8. As we can see, our method have a robust result with the defocus that in the range of 400um, and the results with intensity correction coefficients present best among all. 
\begin{figure}[H]
	\centering
	\includegraphics[scale=0.15]{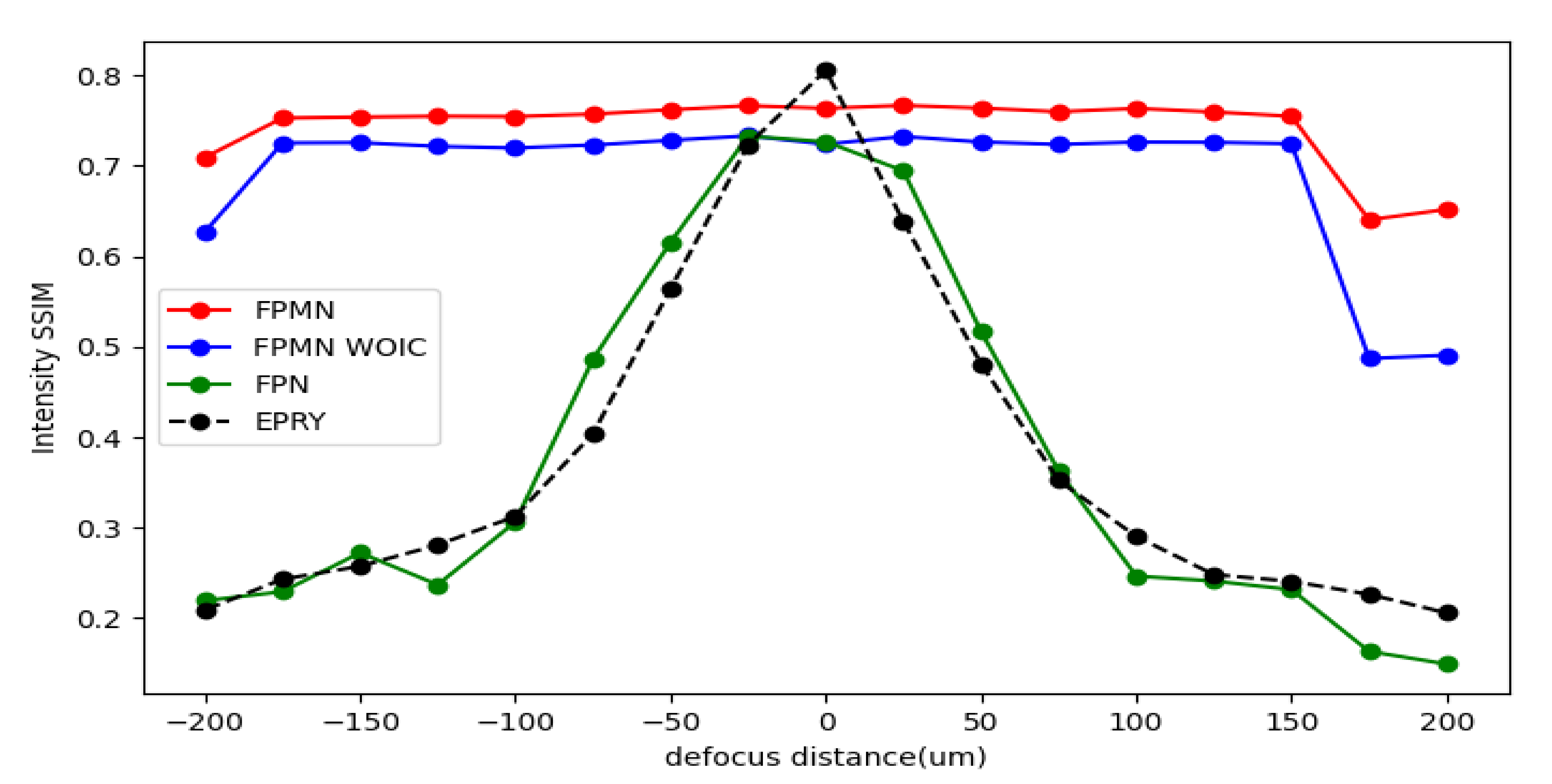}
	\caption{SSIM curve between optim result and GT. Legend FPMN represents the complete optim net. Legend FPMN WOIC represents FPMN without intensity correction coefficients layer. Legend FPN represents FPMN without defocus layer and intensity correction coefficients layer. Legend EPRY represents Embedded pupil function recovery method. }
\end{figure}
\section{Experiment}

\subsection{Experimental results of an USAF resolution target with FPMN}

We bulid a FPM system as shown in Fig.1(b) to acquire LR images. The experimental setup generally followed our simulation parameters, except that the distance between the LED array and the sample plane is set to 92 mm. We use a LED array (CMN, P 2.5, 21$\times$21) for varied-angles illuminations. The wavelength of illuminations is 470 nm and the bandwidth of illuminations is 20nm. A camera (FLIR, BFS-U3-200S6M-C, sensor size 1”, dynamic range 71.89 dB, pixel size 2.4$um$) is used for recording LR images.To demonstrate the feasibility of FPMN, we use an amplitude-only USAF chart as the sample. The USAF chart is placed at the sample plane that adjusted to different defocus distance by a high-precision moving device. 
\begin{figure}[H]
	\centering
	\includegraphics[scale=0.24]{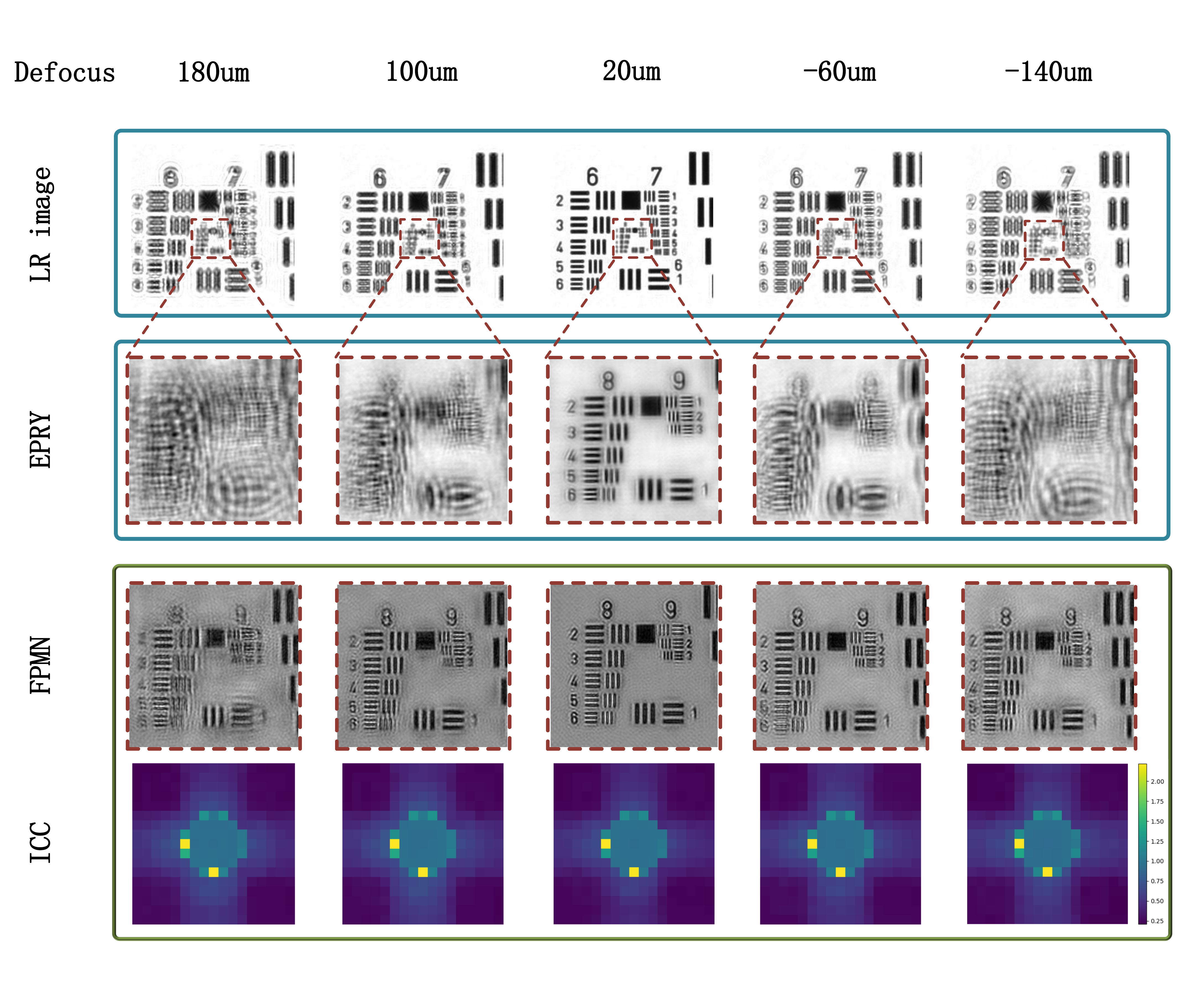}
	\caption{High resolution reconstruction results of EPRY and the proposed
		method with different defocus distances}
\end{figure}
First there is no LED array position deviation in the system, we manually adjust the samples in different defocus positions and compare the results of EPRY and results of our method. The results shown in Fig.9 corresponding to different defocus planes are 180um, 100um, 0um, -60um and -140um respectively. Sequentially, we use the FPMN to realize digital refocusing. The EPRY is used as a comparision experiment. As we can see, when the sample is placed at the focus plane, both the EPRY and FPMN can recover the HR amplitude successfully. But with the defocus distance increase, the performance of EPRY has a degradation, which is consistent with the simulation result shown in Fig.6. In contrast, the reconstructed quality of FPMN can be evidently improved compared with EPRY. 
\begin{figure}[H]
	\centering
	\includegraphics[scale=0.3]{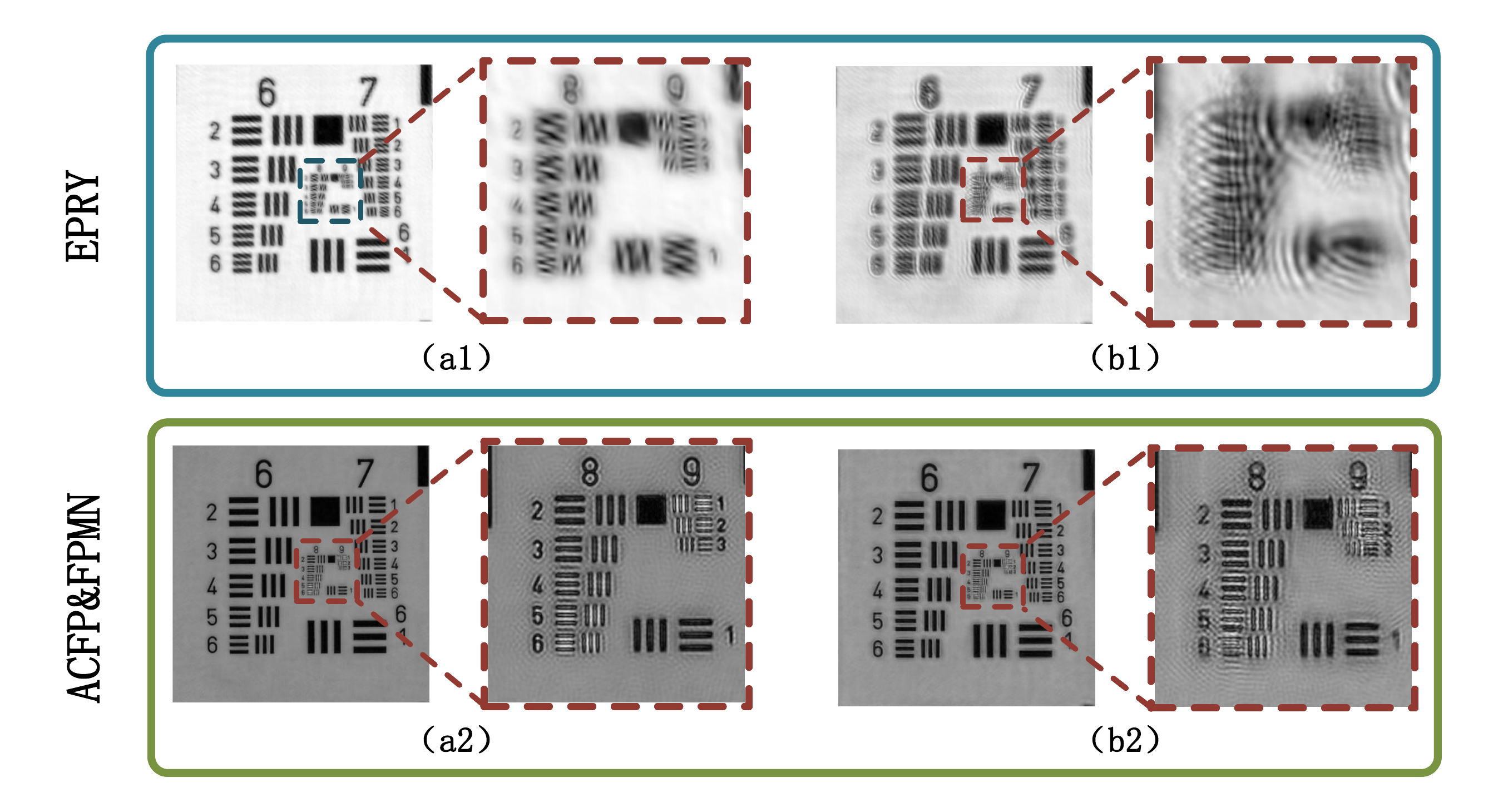}
	\caption{optim results with mixed deviation. (a1)(a2)Results with only LED array position deviation $\Delta x=4mm,\Delta y=4mm,\theta=0.07rad,z=0um$. (b1)(b2)Results with mixed deviation including LED array position deviation and defocus aberration $\Delta x=4mm,\Delta y=4mm,\theta=0.07rad,z=100um$}
\end{figure}
Then we further introduce the LED array position deviation into the system. The results are shown in Fig.10. When the mixed deviation of FPM system exists, the reconstruction results of EPRY has been seriously distorted, but the reconstruction results of FPMN still maintain a high quality with ACFP. These results directly prove the effectiveness of FPMN and ACFP. 

\subsection{Experimental results of biological sample with FPMN}

Generally, a biological sample is in a 3D distribution form, the non-planar distribution characteristic will cause different defocus distance between different subregions. We usually assume that the biological sample is placed at the focus plane in conventional FPM, which may cause degradation of recovery. In this section, a biological sample Paramecium is used to demonstrate the feasibility of the FPMN. The results are shown in Fig.11.

Because of the non-planar distribution characteristic of the biological sample, different subregions of the sample correspond to different defocus distances.The whole FOV image is divided into several subregions to recover by FPMN.
\begin{figure}[H]
	\centering
	\includegraphics[scale=0.3]{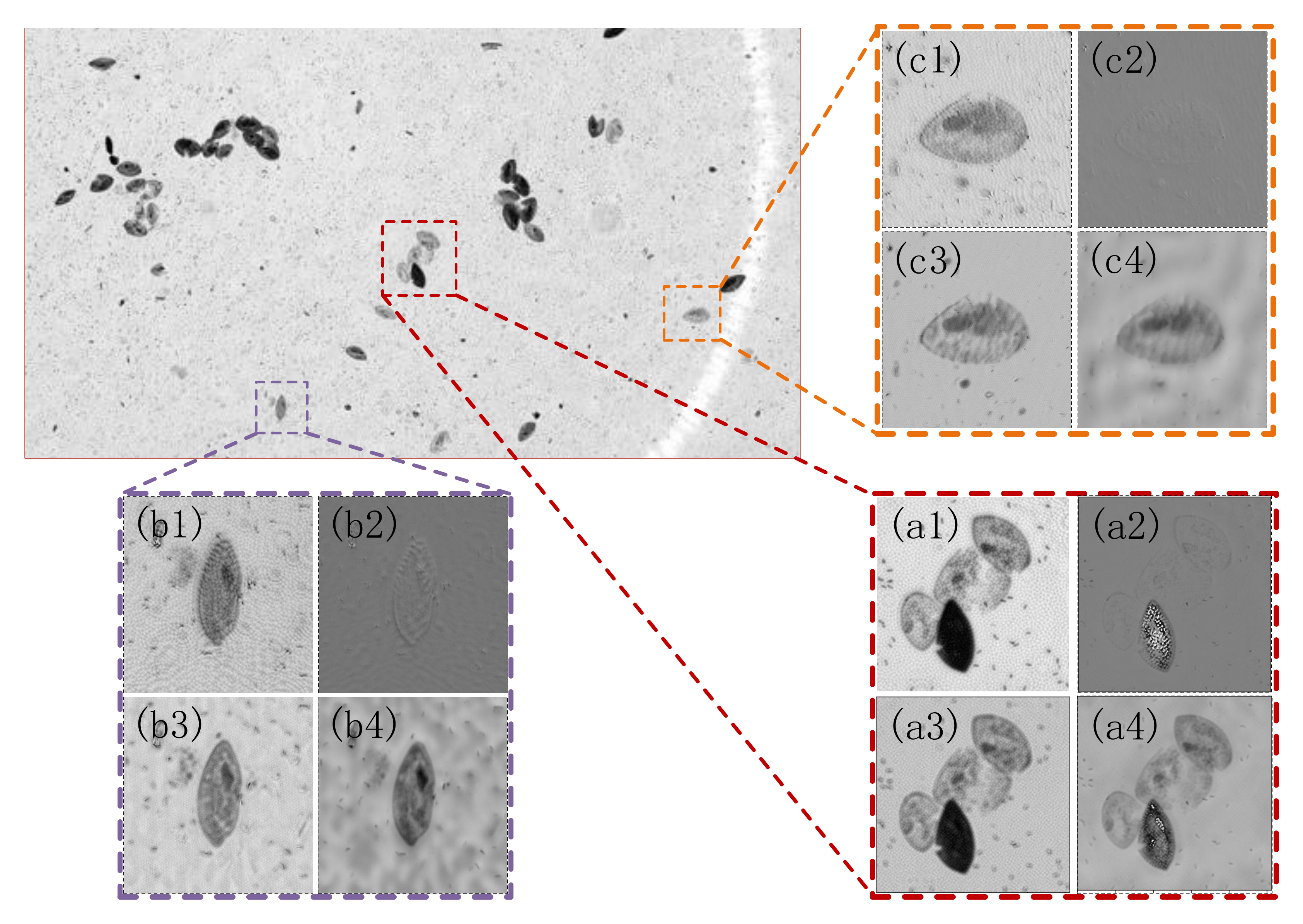}
	\caption{Optim results of biological sample with the EPRY and FPMN. 
		(*1)Reconstructed amplitude with EPRY, (*2)Reconstructed phase with EPRY,(*3)Reconstructed amplitude with FPMN and ACFP. (*4)Reconstructed phase with FPMN and ACFP. }	
\end{figure}
The subregion of Fig.10(a*) is located in the center of the field of view, the subregion of Fig.10(b*) and Fig.10(c*) are located in the edge of the field of view. The object in the edge of the field of view have a relative LED array position deviation, so we use ACFP to obtain the deviation. We compared the reconstruction results of EPRY and our method. As we can see, FPMN with ACFP can improve the reconstructed quality to a certain extent, especially for the phase part, even bring the contrast improve. The results prove the effectiveness of the proposed method. 

\subsection{Different model method of pupil function in FPMN}

In the above experiments, we find that the constraint of prior knowledge can effectively improve the reconstruction quality of the algorithm and reduce the pathologicality of the inverse problem, then achieve the precise decoupling of multiple parameters. 

We used three different modeling methods for pupil function layer in FPMN. The first method is same with fourier obj layer, separates the pupil function layer into real and imaginary parts, The modulation can be express as
\begin{align}
\left\{
\begin{array}{lr}
\mathbf{E}_{nr} =& \hat{\mathbf{O}}_{nr}\cdot \mathbf{P}_{r} - \hat{\mathbf{O}}_{ni}\cdot \mathbf{P}_{i} \\
\mathbf{E}_{ni} =& \hat{\mathbf{O}}_{nr}\cdot \mathbf{P}_{i} + \hat{\mathbf{O}}_{ni}\cdot \mathbf{P}_{r}
\end{array}
\right. 
\end{align}

Considering that coherent transfer function is usually phase modulation for light field in FPM systems. The second method model the pupil function layer as a phase modulation radian value. The modulation can be express as
\begin{align}
\left\{
\begin{array}{lr}
\mathbf{E}_{nr} =& \hat{\mathbf{O}}_{nr}\cdot cos(\mathbf{P}) - \hat{\mathbf{O}}_{ni}\cdot sin(\mathbf{P}) \\
\mathbf{E}_{nr} =& \hat{\mathbf{O}}_{nr}\cdot sin(\mathbf{P}) + \hat{\mathbf{O}}_{ni}\cdot cos(\mathbf{P})
\end{array}
\right.
\end{align}

In order to introduce more priori constraints to FPMN, the pupil function can be fitted with the top ten of the Zenike polynomial\cite{zernike1934diffraction} in optics. In the third modeling method, the pupil function layer uses only the top ten of the zenike polynomial as the parameters to be optimized. which can be express as Eq.(3) and Eq.(4). 
\begin{figure}[H]
	\centering
	\includegraphics[scale=0.16]{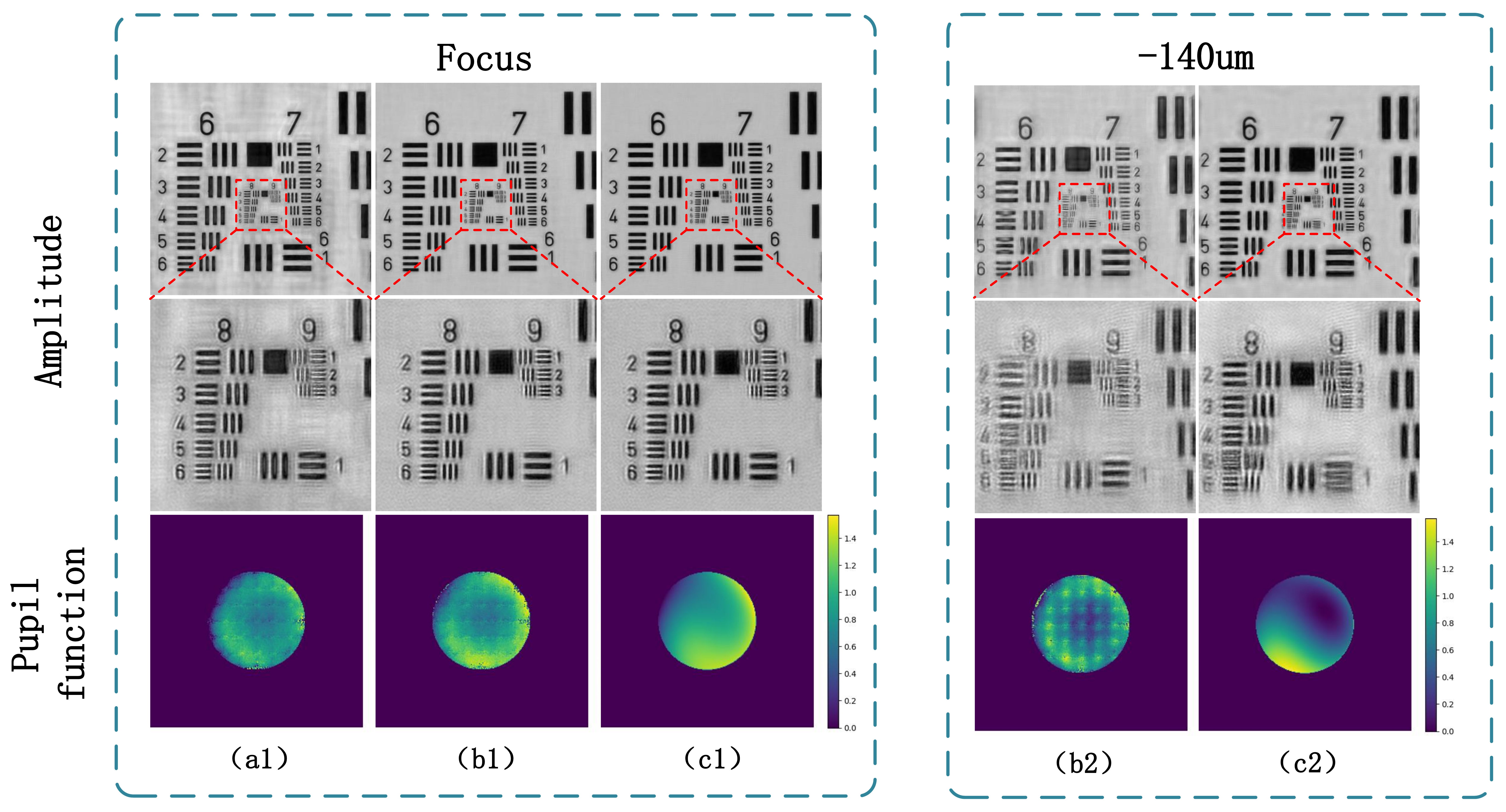}
	\caption{(a1)The optimization results of the modeling method that is same with the fourier obj layer. (b1)(b2)The optimization results of the modeling method with phase modulation radian value. (c1)(c2)The optimization results of the modeling method with zenike polynomial}
\end{figure}
Under three different modeling methods, the results for the sample and pupil function are shown in Fig.12. As we can see, the reconstruction quality of the samples are better under the latter two modeling methods. And when the defocus aberration is introduced, the last modeling method works better. Therefore, the introduction of prior knowledge constraints can improve the decoupling efficiency of multi-parameters

\section{Conclusion and Disscusion}

In this paper, we propose a recovery method for FPM, which uses neural networks to model fourier ptychography imaging processes for optimization. We introduced LED array position deviation correction, defocus aberration correction, and intensity deviation correction into the neuarl network. And we call that neural network FPMN. The LED array position deviation correction method is called ACFP. With ACFP, the FPMN can achieve robustness to mixed deviation, and also can obtain the value of each deviation. 

In order to get better optimization results and faster speed, we should use more priori constraints and more accurate imaging models, as we discussed in this chapter on the way of pupil function modeling. We are considering that using the defocus distance calculated by the geometric relationship in the imaging process\cite{zhang2021fast} as the initial parameter of FPMN, and calculating the fluctuation of LEDs at different angles by diffractive optics knowlodge as the initial parameter of FPMN. That may speed up the optimization\cite{he2015delving} of FPMN and get better decoulping effect. \\ \\
\large \textbf{Funding.} \small National Natural Science Foundation of China (61735003, 61805011). Funding of foundation enhancement program under Grant (2021-JCJQ-JJ-0823) \\ \\
\large \textbf{Disclosures.} \small The authors declare no conflicts of interest.\\ \\
\large \textbf{Data availability.} \small Data underlying the results presented in this paper are not publicly available at this time but may be obtained from the authors upon reasonable request.

\bibliographystyle{unsrt}
\bibliography{reference}

\end{document}